# State-dependent changes of connectivity patterns and functional brain network topology in Autism Spectrum Disorder

Pablo Barttfeld[1], Bruno Wicker[1,2], Sebastián Cukier[1,4], Silvana Navarta[1], Sergio Lew[3], Ramón Leiguarda[4] and Mariano Sigman[1]

1- Laboratory of Integrative Neuroscience, Physics Department, FCEyN UBA and IFIBA, Conicet; Pabellón 1, Ciudad Universitaria, 1428 Buenos Aires, Argentina.

2- Institut de Neurosciences de la Timone, CNRS UMR 7289 & Aix-Marseille University, Marseille, France.

3- Instituto de Ingeniería Biomédica, Facultad de Ingeniería, Universidad de Buenos Aires, Argentina.

4- FLENI, Buenos Aires, Argentina.

e-mail: pablob@df.uba.ar

**Abstract**

Anatomical and functional brain studies have converged to the hypothesis that Autism Spectrum Disorders (ASD) are associated with atypical connectivity. Using a modified resting-state paradigm to drive subjects' attention, we provide evidence of a very marked interaction between ASD brain functional connectivity and cognitive state. We show that functional connectivity changes in opposite ways in ASD and typicals as attention shifts from external world towards one's body generated information. Furthermore, ASD subject alter more markedly than typicals their connectivity across cognitive states. Using differences in brain connectivity across conditions, we classified ASD subjects at a performance around 80% while classification based on the connectivity patterns in any given cognitive state were close to chance. Connectivity between the Anterior Insula and dorsal-anterior Cingulate Cortex showed the highest classification accuracy and its strength increased with ASD severity. These results pave the path for diagnosis of mental pathologies based on functional brain networks obtained from a library of mental states.

# 1. Introduction

Autism Spectrum Disorders (ASD) refer to neurodevelopmental disorders characterized by poor social communication abilities in combination with repetitive behaviours and restricted interests (APA, 2000). Research in the physiopathology of ASD has largely focused on the identification of structural or functional brain abnormalities. Structural MRI studies have reported abnormal developmental trajectory of brain growth, with evidence of poorly organized white matter (Alexander et al., 2007; Barnea-Goraly, Lotspeich, & Reiss, 2010; Egaas, Courchesne, & Saitoh, 1995; Fletcher et al., 2010; Herbert et al., 2005; McAlonan et al., 2005; Thakkar et al., 2008), atypicalities in gyration and cortical thickness patterns (Hadjikhani, Joseph, Snyder, & Tager-Flusberg, 2006; Hardan, Muddasani, Vemulapalli, Keshavan, & Minshew, 2006), possibly caused by irregular neuronal migration, cortical organization and myelinisation in ASD (Wass, 2011). Functional neuroimaging studies using a range of experimental tasks targeting emotional and social information processing have reported hypoactivation in the fusiform gyrus, the amygdala, the dorsomedial prefrontal cortex, the superior temporal sulcus, and insula (Di Martino et al., 2009; Hadjikhani, Joseph, Snyder, & Tager-Flusberg, 2006). A recent shift in emphasis to investigating dynamic processes of functional brain connectivity has led to a convergence on the hypothesis that ASD is associated with atypical connectivity, producing a system that is ineffective for integrating information. Critically, the study of functional brain networks of ASD and typical subjects in the resting state (i.e. during free thought) showed qualitatively different organizations at the group level, which broadly reflects an excess of local connectivity and a deficit in long-range connectivity, especially along the long distance fronto-posterior axis (Schipul, Keller, & Just, 2011) although an excess in local connectivity (over-connectivity) has also been observed (Barttfeld et al., 2011; Belmonte et al., 2004; Just, Cherkassky, Keller, Kana, & Minshew, 2007; Wass, 2011). At the neuronal level, these findings are supported by data revealing a different ASD phenotype in neurons and axons that make up the brain's communication system (van Kooten et al., 2008; Zikopoulos & Barbas, 2010).

While the differences in brain network connectivity are significant and robust at the group level, the effect is not strong enough to classify robustly whether an individual belongs to the typical or ASD group. In this respect, the main objectives of the present work are: first at the theoretical level, to investigate with a systems level approach whether the difference in large-scale brain network organization between ASD and typical subjects depends on the specific task in which subjects are engaged. Second at the practical level, to investigate whether brain network connectivity differences in specific mental states may be more informative to distinguish the organization of the typical and ASD mind.

We measured functional brain connectivity in three different mental states, varying the focus of attention on internal stimulus (focus on respiration, Interoceptive state), external auditory stimulus (oddball task, Exteroceptive state), or having the subjects lying eye closed in the scanner (mindwandering, Rest state). This approach is motivated by the fact that paying attention to an external or internal stimuli is equivalent in term of cognitive demand but qualitatively different regarding the origin of the stimulus to pay attention to, and by the observation that kids with ASD have a heightened interoception (Santos et al., 2010) and have a greater reliance in proprioception which correlates with the impairments in social function and imitation (Haswell, Izawa, Dowell, Mostofsky, & Shadmehr, 2009). Our results show a very marked interaction between group of subjects (ASD or typical) and cognitive state. Network changes between groups in the Interoceptive and Exteroceptive states showed opposite effects, revealing that inferences about connectivity differences in ASD should be state-dependent. Besides, within group brain connectivity differences across cognitive states are more pronounced in the ASD than in the typical group, with brain connectivity in the ASD group being more strongly affected by cognitive state than in the typical group. Furthermore, while network measures are inefficient to decode whether a subject belongs to the ASD or typical group, comparing how network parameters change with cognitive state achieves a significant decoding performance, suggesting that within-group functional change across cognitive states is a better marker of ASD than between-group changes in a given cognitive state.

**2. Material and Methods.**

*2.1 Participants*

Two groups of subjects took part in this study. The ASD group included 12 individuals with high-functioning autism or Asperger's Syndrome (9 men and 3 women; mean age = 23.7, std =7,13). The participants with ASD were provisionally included in the study if they had received a diagnosis of autism or Asperger's Syndrome from a psychiatrist or licensed clinical psychologist. Actual participation required that this diagnosis be recently confirmed, with each having met the criteria for ASD within the past 3 years on the basis of the revised fourth edition of the Diagnostic and Statistical Manual of Mental Disorders and on the score on the Autistic Diagnostic Observation Schedule-Generic (Lord et al., 2000). IQs were measured with the third edition of the Wechsler Adult Intelligence Scale and ranged from 85 to 121 (mean=101.33, SD=13.79). At the time of testing, no ASD subject had known associated medical disorders. ASD participants were matched to a group of 12 typically developing individuals (8 men and 4 women; mean age= 28,83; std= 5,00; IQ: mean=108.91 SD=12.62). Participants in the typical group had a history free of psychiatric disorders.

*2.2 fMRI acquisition.*

Functional images were acquired on GE Hdx 3T with a conventional 8 channels head coil. Twenty four axial slices (5 mm thick) were acquired parallel to the plane connecting the anterior and posterior commissures and covering the whole brain (TR = 2000 ms, TE = 35 ms, flip angle = 90). To aid in the localization of functional data, high-resolution images (3D Fast SPGR-IR, inversion time 700 mm; FA=15; FOV=192x256x256 mm; matrix 512x512x168; slice thickness 1.1 mm) were also acquired.

Subjects underwent three functional runs lasting 7 minutes 22 seconds each. During all runs a series of tones were presented at very low volume within the noise of the scanner. The duration of each tone was 200 ms and the inter-tone interval was 400 ms. The pitch of the majority (94%) of the tones of the sequence was 400 Hz. The remaining tones (referred as oddballs) had a slightly higher pitch (410 hz), and were presented, on average, every 15 tones. In the Rest state run, subjects were instructed to keep their eyes closed, and avoid moving and falling asleep. In the Interoceptive state run, subjects were instructed to focus on their respiration cycle, perceiving the air flowing in and out. In the Exteroceptive state run, participants were asked to focus on the sequence of sounds and identify the oddballs. At the end of the experiment we asked subjects whether they heard the tones in the Rest and Interoceptive runs. None of the subjects (0 out of 24, including participants in both groups) reported noticing the tones in the Rest or Interoceptive state, which indicates that the tones were well camouflaged within the noise of the scanner in absence of voluntary directed attention to them. A strict measure of audibility of the tones was not necessary for the purpose of this study since our aim was to broadly direct attention endogenously or exogenously.

*2.3 Data processing and analysis*

Functional data were preprocessed using statistical parametric mapping software (SPM5; http://fil.ion.ucl.ac.uk/spm). The first 4 volumes of each run were discarded to allow for longitudinal relaxation time equilibration. EPI images from all sessions were slice-time corrected and aligned to the first volume of the first session of scanning to correct for head movement between scans. There was no excessive motion in any of the scans (lower than 3 mm). A mean image was created using the realigned volumes. T1-weighted structural images were first co-registered to the mean EPI image of each participant. Normalization parameters between the co-registered T1 and the standard MNI T1 template were then calculated, and applied to the anatomy and all EPI volumes. Data were then smoothed using an 8 mm full-width-at-half-maximum isotropic Gaussian kernel to accommodate for inter-subject differences in anatomy.

One hundred-sixty previously published regions defining 6 functional networks (Fronto-parietal (FP), Cingulo-Opercular (OP), Default Brain Network (DEF, Occipital (OC), Sensorimotor (SE) and Cerebellum (CER) were used to build spherical ROIs defined as the set of voxels contained in a 5-mm sphere around a coordinate (Dosenbach et al., 2010). The mean time course in each ROI was extracted by averaging the time courses of all of the voxels contained in the ROI (http://marsbar.sourceforge.net). For each ROI, a time series was extracted separately for each individual and each experimental condition. These regional fMRI time series were then used to construct a 160-node functional connectivity network for each subject and condition. We used wavelet analysis to construct correlation matrices from the time series (Supekar, Menon, Rubin, Musen, & Greicius, 2008). We followed the procedures exactly as described by Supekar and collaborators: We applied a maximum overlap discrete wavelet transform (MODWT) to each of the time series to obtain the contributing signal in the following three frequency components: scale 1 (0.13 to 0.25 Hz), scale 2 (0.06 to 0.12 Hz), and scale 3 (0.01 to 0.05 Hz). All subsequent analysis was done based on the scale 3 component, whose frequency lies in the range of slow frequency correlations of the Default network (Fox et al., 2005; Raichle, 2009). To account for the relatively small number (220) of data points per time series for low frequency correlation analysis, the vector representing the time series beyond its boundaries (0 and 220) was assumed to be a symmetric reflection of itself (Supekar, Menon, Rubin, Musen, & Greicius, 2008). The resulting connectivity matrices describe frequency-dependent correlations, a measure of functional connectivity, between spatially-distinct brain regions.

We also conducted a seed analysis (Fox et al., 2005), with the five most relevant ROIs for the classification analysis (i.e. those ROIS that most contributed to classify the subjects). These ROIS were -in descending relevance- from CO network: Anterior Insula [38, 21, -1], Dorsal Anterior Cingulate [9, 20, 34], Anterior Prefrontal Cortex [27, 49, 26], Basal Ganglia [-6, 17, 34], Medial Frontal Cortex [0, 15, 45]; from FP network: Intraparietal Sulcus [32, -59, 41], Dorsal Frontal Cortex [44, 8, 34], Dorsal Frontal Cortex [40, 17, 40], Inferior Parietal Lobule [44, -52, 47], Dorsal Frontal Cortex [-42, 7, 36]; from DEF network: Precuneus [9, -43, 25], left Inferior Temporal Cortex [-61, -41, -2], Ventromedial Prefrontal Cortex [9, 51, 16], Angular Gyrus [51, -59, 34], right Inferior Temporal Cortex [52, -15, -13].

To obtain a brain volume containing the correlation of every voxel with the seed ROI, we followed the same methods than for the connectivity analysis, calculating the wavelet-correlation at low frequencies between each seed's time series and all voxel time series for each subject and condition, and remapping the resulting correlation vector into a 3D brain volume of the same size than the normalized images, obtaining five volumes per subject and condition (one per ROI). We averaged these five volumes to get an average volume representing the average connectivity between all voxels in the brain and the five ROIs. Using these correlation volumes we conducted a second-level multiple regression analysis, including IQ and sex as covariables of no interest, to regress each correlation value to the ADOS score of each ASD subjects. To account for multiple comparisons, the resulting statistical images were assessed for cluster-wise significance using a cluster-defined threshold of p=0.005; extent threshold=5 voxels.

*3 Graph Theory metrics*

The connectivity matrix defines a weighted graph where each electrode corresponds to a node and the weight of each link is determined by the wavelets correlation at low frequency. To calculate network measures, functional connectivity matrices were converted to binary undirected matrices by applying a threshold T. We explored a broad range of threshold values of 0.0005 < T < 1, with increments of 0.001- and repeated the full analysis for each value of T. After transforming the functional connectivity matrices to a binary undirected graph, we measured the Degree (K), Path Length (L) and Clustering Coefficient (C) using the BCT toolbox (Sporns & Zwi, 2004). Network visualizations were performed using the Pajek software package (http://pajek.imfm.si/doku.php) using a Fruchterman–Reingold layout algorithm (Fruchterman & Reingold, 1991). Differences between groups were assessed by means of independent ANOVAs with group and threshold (the cutoff to determine whether two ROIs are connected) as independent factors. A broad range of thresholds was used, from 50 to 750, in steps of 50 (excluding the extreme values where networks disaggregate).

*4 Classification analysis*

In order to investigate the interaction between cognitive states and individual ROI connectivity, we conducted a multi-parametric classification analysis using support vector machines (SVM) (Theodoridis, 2009), to classify subjects as typical and ASD. We selected the degree (K) of each individual ROI as features, averaged across all thresholds to avoid the use of an arbitrary one. We chose SVM because they are resilient to over-fitting and allow the extraction of feature weights (Dosenbach et al., 2010; Formisano, De Martino, Bonte, & Goebel, 2008; Norman, Polyn, Detre, & Haxby, 2006) to explore in an unbiased way the ROIs that better characterize ASD changes across conditions. A leave-one-out-cross-validation was used to estimate the significance of the classification performance. This is a frequently used method because it allows the use of most of the data for training (Fukunaga & Hummels, 1989). In this procedure, all samples except one are used to train the SVM. The remaining sample is used to test the decision function derived from the training stage. Each sample is designated only once as test, and the final accuracy of the SVM is calculated averaging the accuracies for all test stages (repeating the analysis for each sample as test sample). A permutation analysis was used to assess statistics of the classification procedure (Golland & Fischl, 2003). The whole process of classification was repeated 10.000 times, randomizing class labels to estimate an empirical distribution of the classifier accuracies under the hypothesis of no discriminability (no actual group separation). P values are estimated as the proportion of accuracies in the null-distribution higher than the observed accuracy.

## 3. Results

For each state *s* (Interoceptive, Exteroceptive or Rest) and participant *p* (24, 12 of each group) we measured a 160*160 connectivity matrix where the matrix entry $C(s,p)_{ij}$ indicates the lag-0 correlation of the average fMRI signal of ROIs *i* and *j* for participant *p* in state *s*. As expected from previous work (Dosenbach et al., 2007; Dosenbach et al., 2010), simple inspection of correlation matrices for all conditions and groups (Figure 1a) reveals that, the ROIs within each functional system form clusters, with closely correlated temporal profiles.

| group | Age | Total IQ | ADOS | | |
|---|---|---|---|---|---|
| | | | Comm. | Soc. Int. | *Total* |
| Control | 28,8 (5,00) | 108,91 (12,62) | - | - | - |
| ASD | 23,7 (7,13) | 101,33 (13,79) | 3,5 | 6,66 | 10,16 |

**Table 1**. Details of ASD and Typical subjects: diagnosis, IQ and ADOS scores.

For each state, we conducted t-tests for each entry of the matrix comparing the ASD and typical group of subjects (Figure 1a). A positive t-value indicates that connectivity increased in ASD compared to typical population. Conversely, a negative t-value indicates that connectivity is greater in the typical than in the ASD population. Figure 1b shows the distribution of absolute t-values to picture an unsigned estimate of change across groups for each cognitive state). The distributions of t-values differences between groups is shifted towards negative values, showing a strong decreased connectivity in ASD compared to typicals in the Exteroceptive condition (mean = -0.5539; std = 0.937; t-value = -94.53, Cimin = -0.5654; Cimax = -0.5424). The same trend is observed in the Rest state condition (mean = -0.0938; std = 1.035; t-value = -14.507; CImin = -0.10; CImax = -0.08). By contrast, an opposite effect is observed in the Introspective state condition, where there is a very pronounced increase of connectivity in the ASD compared to typical group (mean = 0.4837; std = 1.059; t-value = 73.053, Cimin = 0.4707; Cimax = 0.496). This shows that, averaging across all pairs of regions, connectivity increases in the ASD group for the Interoceptive condition and de-creases for the Exteroceptive state condition. To better quantify this we conducted independent mixed ANOVAs with group as a between subjects factor, and type of connection (21 possible symmetrical combinations between and within functional the 6 functional systems) as within factor. Full results of the ANOVA are listed on table S1. The effect of connection type was significant for the three cognitive states, confirming the functional segregation of the 6 functional systems and an inhomogeneity in their specific pattern of connection (Dosenbach et al., 2010). In the Exteroceptive and Interocepctive states we observed a significant effect of group with no interaction between both factors: Effect of group in the Exteroceptive ($F(1, 231) = 15.55$; $p < 0.001$; mean typical = 0.4089; mean ASD = 0.3565), and Interoceptive ($F(1,231) =$ 9.72; p <0.01; mean typical = 0.3772; mean ASD = 0.4205). Instead, in the Rest state the effect of group was non significant (F(1,231) = 0,58; p > 0,1; mean typical = 0.3741; mean ASD = 0.3664), with no interaction between type of connection and group. These results reveal that, as seemed apparent from an inspection of Figure 1, average functional brain connectivity decreases in the Exteroceptive state condition and increases in the Interoceptive state condition in ASD compared to typicals. In the Rest state condition the excess and deficit of connectivity balance.

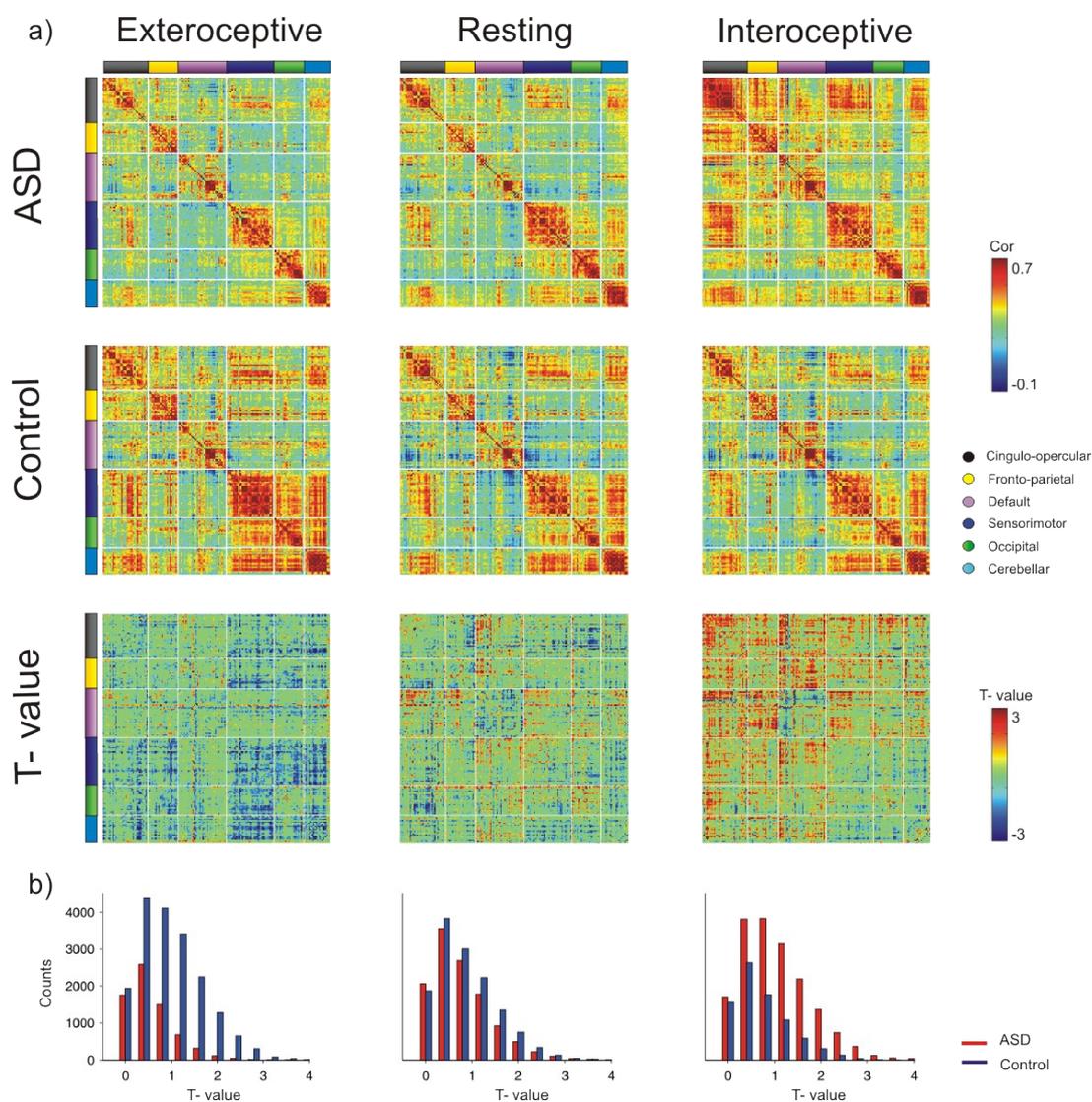

**Figure 1.** Networks connectivity matrices. a) Averaged correlation matrices, for groups and conditions. Bottom raw shows t-values for test-t between groups. b) T-value distributions for ASD (red) and Typicals (Blue).

The previous analysis investigated how brain connectivity networks varied across groups for each cognitive state. Another possible way of analysing our factorial dataset is to investigate how networks vary within states for each group. Simple inspection of the connectivity matrices (Figure 1, Supplementary Figure 2) suggests that changes in connectivity matrices are more pronounced across states in ASD subjects than in typicals. To quantify this observation we conducted within group t-tests for each entry of the matrix comparing connectivity between Exteroceptive and Interoceptive state conditions (Supplementary Figure 2). A positive t-value indicates that connectivity increased in the Interoceptive state compared to the Exteroceptive state condition. Conversely, a negative t-value indicates that connectivity is greater in the Exteroceptive than in the Interoceptive state condition. The distributions of t-values differences for the typical group is almost centered at zero (mean = 0.01; std = 0.734; t-value = -3.43, Cimin = 0.0068; Cimax = 0.024), showing that there is no net change in connectivity across conditions. On the contrary, the distribution of t-values for the ASD group is strongly shifted to positive values (mean = 0.4837; std = 1.059; t-value = 112.19, Cimin = 0.5575; Cimax = 0.5773). This shows that connectivity in ASD subjects strongly fluctuates in different cognitive states while in the typical subjects it shows modest fluctuations and remains relatively stable. The systems showing greater variability in connectivity across states in the ASD groups were the CO, FP and DEF (Supplementary Figure 2).

To investigate the impact of changes in connectivity in network topology, we constructed functional networks assigning one node for each ROI and considering a link between nodes if connectivity exceeded a fixed threshold. We used an arbitrary threshold of 0.55 and embedded the network in the two-dimensional plane using a force-directed algorithm of energy minimization (Fruchterman & Reingold, 1991). This is merely for visualization purposes; the results described here were robusts within a wide range of thresholds and statistical analysis used a varying threshold as an independent factor (see Supplementary Figure S3). The embedded functional networks revealed patterns consistent with Figure 1 and allow to further zoom in specific topological changes between groups and states. Networks of both groups are quite similar in the Rest state condition, with the exception of the Default system whose nodes are quite apart in the ASD network but form a compact cluster in the Typical network (Figure 2a, pink dots, centre). This is consistent with previous observations identifying a main change in the connectivity of the default system between ASD and typicals in the resting state (Belmonte et al., 2004; Courchesne, Redcay, Morgan, & Kennedy, 2005; Just, Cherkassky, Keller, Kana, & Minshew, 2007; Kennedy & Courchesne, 2008; Monk et al., 2009; Wass, 2011; Weng, 2010). Networks obtained in the Exteroceptive state condition show more pronounced topological differences across both groups: The typical network is more packed and clustered, suggesting a lower diameter of the entire network (Figure 2a, left). by contrast, the ASD network is more compact in the Interoceptive state condition. We observe that the Cingulo-Opercular (black dots) system is more tightly packed (self-connected) for the ASD group, and closely connected to the Fronto-parietal (yellow) and Default (pink) systems (Figure 2a, right).

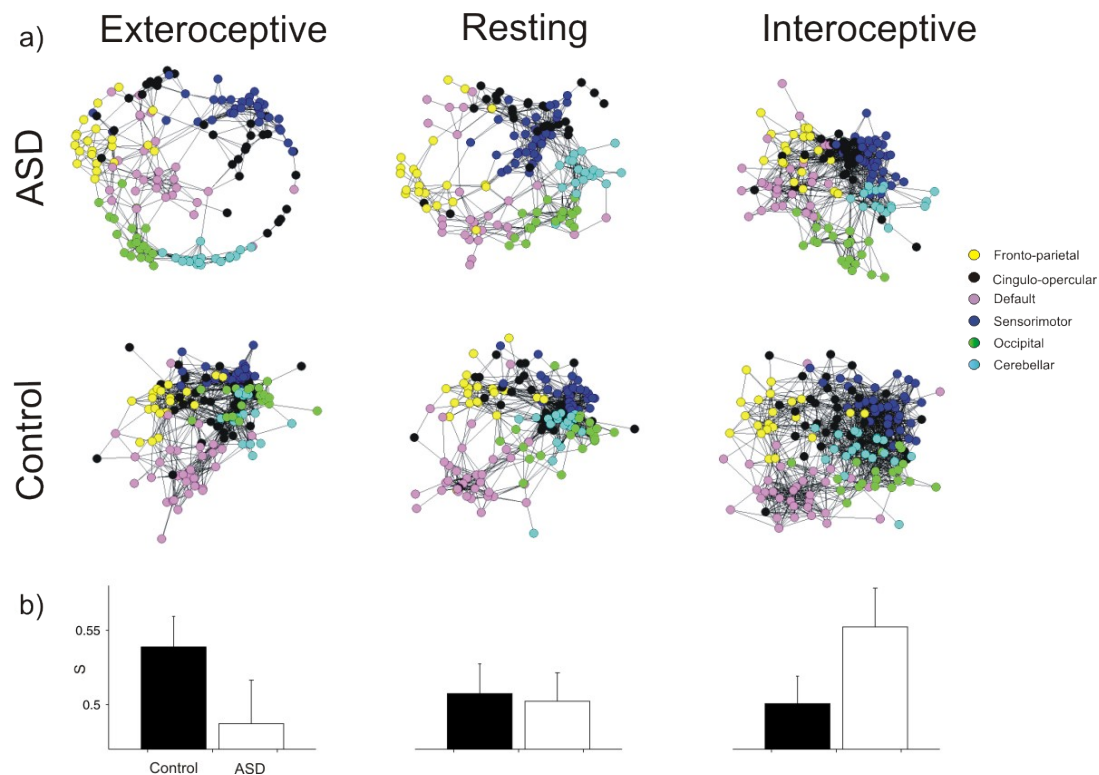

**Figure 2.** Organization of functional brain networks and network metrics. a) Two-dimensional projections of the networks for ASD and typical subjects for all conditions b) Comparison of S=C/L which estimates the small-worldness of a network for typical and ASD. ASD subjects present lower S in Exteroceptive, almost equal S in Resting and higher S in Interoceptive.

To quantify these observations we performed independent ANOVA analyses for the Path Length (L) and the Clustering Coefficient (C) (Gallos, Song, Havlin, & Makse, 2007; Sporns & Zwi, 2004) and for each states with group and threshold (the cutoff to determine whether two ROIs are connected) as independent factors. Threshold effect was significant for all conditions, while interaction between group and threshold was never significant. This means that threshold selection did not change the pattern observed across groups. Comparisons between typical and ASD revealed an effect of group on C (higher in ASD) and L (lower in ASD) in the Interoceptive state condition (C: $F(1,330) = 4,60$; $p < 0,05$; L: $F(1,330) = 8,52$; $p < 0,01$). Comparisons between groups also revealed a main effect of group in the Exteroceptive state condition for C (lower in ASD) and L (higher in ASD) (C: $F(1,330) = 6,60$; $p < 0,01$; L: $F(1,330) = 8,73$; $p < 0,01$). On the contrary, there was no effect either for C nor for L in the Resting state condition (C: $F(1,330) = 0,17$; $p > 0,1$; L: $F(1,330) = 0,02$; $p > 0,5$). The combined changes of path length and clustering can be combined in the quotient S=C/L (Figure 2b), which indicates how closely the network is organized as a Small-World (The higher S, the more small-world organization). Small-world refers to an ubiquitous present topological network which has a relatively short (compared to random networks) characteristic path length (L) and high clustering coefficient (C)(Watts & Strogatz, 1998). Our results show that S increases for the typical compared to the ASD group in the Exteroceptive state condition ($F(1,330) = 10,01$; $p < 0,01$). Instead, S is higher for the ASD than the typical group in the Interoceptive state condition ($F(1,330) = 17,4$; $p < 0,0001$). There are no significant differences in the Resting state condition ($F(1,330) = 0,27$; $p > 0,5$) (Figure 2b).

The next aim was to investigate which aspects of connectivity are more informative to classify subjects as belonging to the typical or ASD group. For classification purposes, we used a support vector machine (SVM), an algorithm from the Machine Learning field widely used in the classification of neuroimages (Ecker, Marquand et al., 2010; Kloppel et al., 2011; Theodoridis, 2009; Vapnik, 1998). We investigated the classification power of functional connectivity using the degree of each ROI, which broadly estimates mean connectivity, as the main classification feature. Results, based on a leave-one-out procedure (see Experimental procedures) showed non significant classification rate for all cognitive states: Resting = 0.50; $p > 0.1$; Interoceptive = 0.50; $p > 0.1$. Exteroceptive = 0.45; $p > 0.1$). Based on our previous analysis which showing a very marked interaction between group and cognitive state, we reasoned that instead of considering the connectivity within one cognitive state a better classification feature might be the difference in connectivity between the Exteroceptive and Interoceptive states. When the difference in connectivity between these two states was used as a feature, we indeed observed a significantly better classification (0.75; $p < 0.01$). Other differences did not show any significant results (Exteroceptive – Resting = 0.62; $p > 0.1$.

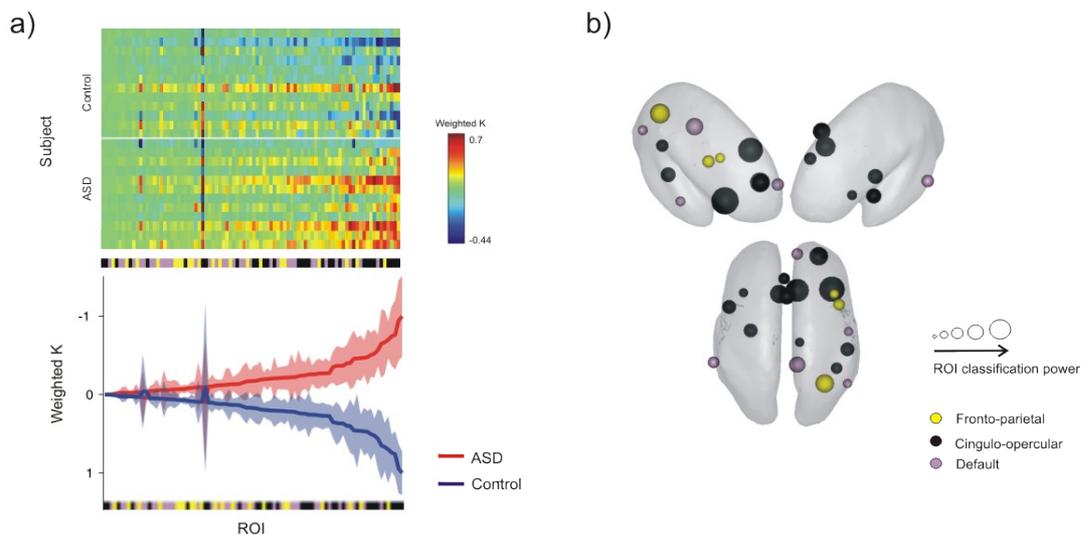

**Figure 3.** Classification analysis. a) Weighted degree for all subjects, sorted by the difference between groups. b) Averaged across subjects per group showing the incremental difference between degree between groups. Shadow represents s.e.m. C) Top 20 ROIs for the classification analysis.

Interoceptive – Resting = 0.50; p > 0.1). A direct comparison of variability in connectivity across states for the ASD condition was mostly concentrated in the Fronto-parietal (FP), Cingulo-Opercular (CO) and Default (DEF) (Supplementary Figure 2). These are particularly relevant brain structures since they account for a segregation of executive control in adaptive control (FP) and stable set-maintenance (CO) functions (Dosenbach et al., 2007). They have also been proposed to play a key role in psychopathology (Menon, 2011). As expected, when we restricted to the CO-FP-DEF set of ROIs, performance in the leave-one out procedure achieved higher scores, reaching performance above 90% for the Interoceptive – Exteroceptive difference (0.9167; p < 0.0001) (See Table S2).

An interesting aspect of SVM analysis is that it allows to measure the weight with which each ROI contributes to the classification (Figure 3). To visualize the relative contribution of different systems and ROIs to classification, we sorted the value of all features weighted by their classification power. This analysis showed that the majority of ROIs that better separate ASD from typicals belong to the cingulo opercular (CO) system. The four regions with the strongest classification power are: right Anterior Insula [38, 21, -1], the Dorsal Anterior Cingulate [9, 20, 34], the right Anterior Prefrontal Cortex [27, 49, 26] and the Basal Ganglia [-6, 17, 34]), which belong to the CO. Within the DEF system, the ROIs with strongest classification power are the Precuneus [9, -43, 25], left Inferior Temporal Cortex [-61, -41, -2], Ventromedial Prefrontal Cortex [9, 51, 16], right Angular Gyrus [51, -59, 34], and right Inferior Temporal Cortex [52, -15, -13]. In the FP system, the ROIs that are within the top 20 contributers to classification are right Intraparietal

Sulcus [32, -59, 41], right Dorsal Frontal Cortex [44, 8, 34 and 40, 17, 40]], and right Inferior Parietal Lobule [44, -52, 47].

The preceding analysis is based on categorical classification, where each participant is assigned to the typical or ASD group. Although significant, this classification is based on a relatively small sample size (a total of 24 subjects, 12 belonging to each category). A more taxing way to investigate the impact of specific connectivity patterns and their dependence with cognitive state on ASD is to investigate progressive changes in connectivity with a continuous progression of ASD severity. If the observed differences in connectivity between groups truly characterize ASD, these differences should also progress according to the severity of ASD. ASD severity is measured through ADOS score which varied from 7 (the minimal value for ASD diagnosis) to 16 in our clinical population. We examined how the connectivity between ROIs and the rest of the brain covaried with ADOS.

To be able to visualize the topography of this covariation without presenting an overwhelming amount of data we calculated the correlation between each of the three main systems (FP, CO and DEF) that were previously identified by the classification and variance analysis as the most sensitive systems to detect differences between typicals and ASD and the rest of the brain. For each of these three systems, we first calculated the five ROIs which ranked higher in their classification power. We then ran an independent seed analysis measuring the average correlation of the five ROIs to all voxels in the brain. Finally, we averaged the five correlation brain volumes of each system to obtain a single volume per subject and system. In a second (group) level of analysis, the resulting correlation volumes calculated for each individual and system were submitted to a linear model using ADOS score as a regressor, and Intelligence Quotient (IQ) and sex as potential regressors of no interest which could account for the variance. To account for multiple comparisons, the resulting statistical images were assessed for cluster-wise significance using a cluster-defined threshold of p=0.005; extent threshold=5 voxels. We observed a strong dependence of connectivity with ADOS in the Resting and Interoceptive state conditions (Figure 4).A full detail of the correlation maps is presented in Table S4 We refer below to the most relevant findings, but a full detail of the correlation maps is presented in Table S4. First we note that none of the correlations was significant in the Exteroceptive state condition. This finding is somewhat puzzling since we had observed that at the group level connectivity decreased for the ASD compared to typicals. This result shows that this effect is not smoothly graded with the severity of the disease. Second, we observe a main trend in regions showing negative covariation (i.e. whose connectivity decreases with ADOS). Negative covariations are only observed in the Rest state condition. They are largely localized within the posterior brain in the parietal, occipital and temporal regions, which tend to be more disconnected from the three main systems (CO, FP and DEF). Instead, regions showing positive correlations (i.e. whose connectivity increases with ADOS) are mainly observed in the Interoceptive state condition and largely localized in the more dorsal regions of the brain.

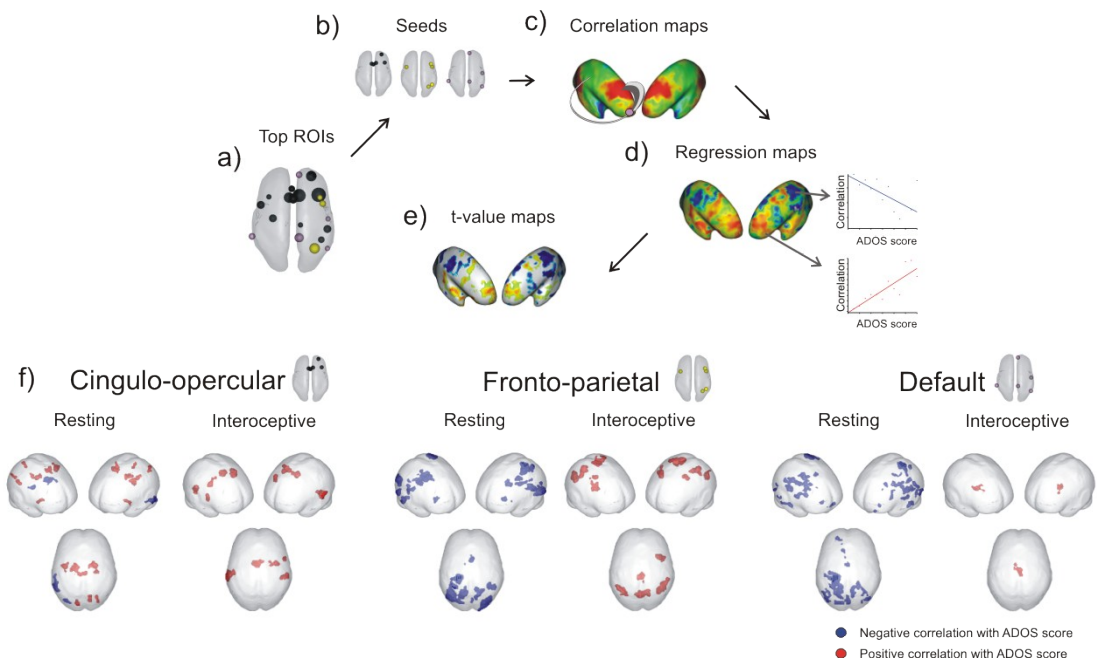

**Figure 4.** Projection of statistically significant clusters for positive (red) and negative (blue) correlation between ADOS and connectivity. All negative relations between ADOS and connectivity are found in Neutral condition. Most of the positive relations between ADOS and connectivity are found in Interoceptive conditions. There are no significant correlation between ADOS and Exteroceptive condition.

## 4. Discussion

The purpose of this study was to characterize and compare large-scale functional brain connectivity networks in ASD and typical subjects in three different cognitive states. The main novelty of our work lies in demonstrating that connectivity changes between ASD and typical populations are state dependent. Beyond the specific functional properties of the brain areas involved, which we review below, our results show a consistent global trend in the ASD relative to the typical population: brain network connectivity increases when attention is directed towards an internal stimulus

(Interoceptive state) and decreases when attention is directed to an external stimulus (Exteroceptive state). In the Rest state, our data replicate previous findings suggesting a decrease in the intrinsic connectivity of the default system in ASD (Belmonte et al., 2004; Just, Cherkassky, Keller, Kana, & Minshew, 2007; Kennedy & Courchesne, 2008; Monk et al., 2009; Wass, 2011; Weng, 2010). In addition, classification analysis revealed that changes in brain networks connectivity in different mental states are more informative than direct comparison of brain networks between populations in a given state to distinguish typical and ASD subjects.

Neuroanatomical brain abnormalities in psychiatric pathologies have previously been investigated as potential classifiers aiming to assist or refine diagnosis (Ecker, Marquand et al., 2010; Kloppel et al., 2011; Kloppel et al., 2008). Most studies using a classification approach to explore specific neuropathological underpinnings and brain-based biomarkers of ASD have used sets of morphological data parameters such as cortical thickness or gray matter volume (Ecker, Marquand et al., 2010; Ecker, Rocha-Rego et al., 2010; Uddin et al., 2011). To date, no published studies have used functional imaging data to identify precisely which brain regions can be used to discriminate individuals with ASD from typically developing individuals. Our study is the first to apply analytic SVM approach to functional imaging data and large scale network modelisation. We show that the properties of functional brain connectivity networks in any given cognitive state are poor classifiers to separate ASD and typical populations. Quite timely, our results thus reveal that the exploration of resting state within a library of other mental states contains much more information to classify ASD and typical subjects. This may constitute a powerful methodology beyond the specific domain of ASD.

The linear classifier method we used has the advantage to enable the identification of which specific brain regions contribute the most to the classification process (Figure 3). Interestingly, all six regions showing the highest classification accuracy in our study -anterior insula (AI), dorsal Anterior Cingulate Cortex (dACC), anterior Prefrontal Cortex (aPFC), Intra-Parietals Sulcus (IPS), Middle Frontal Cortex (mFC), and Precuneus (PC) - are relevant to the ASD pathology.

In our data, the Anterior Insula (AI) showed the greatest classification accuracy and appears as a key region showing differences between Exteroceptive and Interoceptive attentional states. It is also one of the brain regions showing the strongest correlation with ADOS score. Several independent sources of evidence predict that the AI would play a key role in this classification process. First, AI is situated at the interface between the cognitive, homeostatic and affective systems of the human brain, providing a link between stimulus-driven processing and brain regions involved in monitoring the internal milieu and interoceptive awareness of physiological changes in the body

(Craig, 2002, 2009, 2010, 2011; Menon, 2011). Moreover, Critchley (2004) provided evidence that there are strong links between the right AI, perception of one's own bodily state, and the experience of emotion. The AI would be part of a "salience network" that mediates dynamic interactions between externally oriented attention and internally oriented self-related processes, and serves to integrate sensory data with visceral, autonomic, and hedonic information (Sridharan, Levitin, & Menon, 2008; Uddin & Menon, 2009). Second, it has been shown to be involved in social cognition and emotion processing, through integration of interoceptive informations and body awareness (Lamm & Singer, 2010; Straube & Miltner, 2011). Interestingly, interoceptive representation has been suggested to modulate many skills known to be compromised in ASD, such as motivational behaviour (Craig, 2002), empathy (Lamm, Decety, & Singer, 2011; Lamm & Singer, 2010), processing of basic emotions (Wicker et al., 2003) and social skills such as theory of mind (Bird et al., 2010). In concordance, AI has been reported as significantly hypo-activated and underconnected in ASD during social and empathic tasks (Di Martino et al., 2009; Silani et al., 2008). Third, at the anatomical level, the AI is among the few brain regions (along with the dACC, our second most predictive ROI) containing Von Economo or "spindle" neurons, thought to be unique to higher primates (Nimchinsky et al., 1999) and whose abnormal development may cause the social disabilities characteristic of ASD (Allman, Watson, Tetreault, & Hakeem, 2005; Frith, 2001; Santos et al., 2010).

Craig and collaborators have postulated that the AI is an evolutionary specialization in primates that is tailored to integrate a map of internal bodily states with motivational drives generated in the dACC (see Craig, 2009 for review). Interestingly, the dACC ranks second in classification accuracy and our results demonstrate specifically that dACC and AI connectivity (both within the cingulo-opercular system) show a very marked interaction with condition and cognitive state: their connectivity increases in ASD compared to typical population in the Interoceptive state and decreases in ASD compared to typical population in the Exteroceptive state. Consistent with this, hyper-activation of the dACC to social targets has been reported to predict the severity of social impairments in ASD subjects (Dichter, Felder, & Bodfish, 2009). Similarly, our results are inline with the observed specific increase in functional connectivity between striatal subregions –another Cingulo-opercular ROI– and insula (Di Martino et al., 2011) which in our study is only observed in the Interoceptive state. AI and dACC are believed to constitute the main nodes of the Cingulo-opercular system, and to integrate cognitive, homeostatic and emotional information (Menon, 2011; Menon & Uddin, 2010). Altogether, previous data consistently show that interaction of the AI with other brain systems such as the dACC plays a key role in the mediation of interoceptive and exteroceptive states. The function of this 'salience network' is to identify the most homeostatically relevant among several internal and extrapersonal stimuli in order to guide behavior (Seeley et al., 2007). Our findings suggest that this "salience network" may be affected by ASD condition.

The right aPFC ranked third in our classification analysis. This region has been related to several functions including explicit processing of internal states and the introspective evaluation of one's own thoughts and feelings (Fleming, Weil, Nagy, Dolan, & Rees, 2010). Anterior PFC functioning and structure are also known to be altered in ASD, with evidence of dysfunctionning during executive functioning (Kawakubo et al., 2009), mentalising (Dumontheil, Burgess, & Blakemore, 2008; Schmitz et al., 2006) and self representation (Lombardo et al., 2010). Structurally, the aPFC is characterized by an increased amount of gray matter in ASD (Lombardo et al., 2012). The dorsomedial Prefrontal Cortex (mPFC) has been repeatedly reported as underactivated in ASD, particularly during tasks requiring attribution of mental states or social information processing (Castelli, Frith, Happe, & Frith, 2002). Furthermore, abnormal effective connectivity has been reported between the mPFC and the right lateral anterior prefrontal cortex in a task involving explicit emotional processing in ASD (Wicker et al., 2008). Abnormal development of the medial prefrontal cortex has been proposed in ASD, with evidence of abnormal local over-connectivity and long distance disconnection (Courchesne, Redcay, Morgan, & Kennedy, 2005; Zikopoulos & Barbas, 2010). The mPFC is also a key structure of the default mode network, along with the IPS and the precuneus, two brain regions ranking high in the classification analysis. Abnormal functioning of these structures was reported in several resting state studies in ASD (Kana, Keller, Cherkassky, Minshew, & Just, 2009; Monk et al., 2009).

Recent mathematical efforts have established connecting bridges between connectivity measures and functional properties of the emergent network (Gallos, Song, Havlin, & Makse, 2007; Sporns & Zwi, 2004) and the main relevance of brain connectivity patterns lies in their implications for global function in terms of information transfer and segregation between regions. While at this stage conclusions based on graph metrics are purely speculative, our findings suggest that the ASD functional connectivity brain networks largely vary across conditions: when subjects are asked to focus attention to external stimuli, the associated brain connectivity network reveals sub-optimal metrics, suggesting that ASD networks are badly suited for this kind of information processing. When attention is focused on internally generated stimuli, ASD brain networks improve their metrics –even surpassing those of typicals'- suggesting that ASD networks may be better tuned for interoception. Although the processing of an interoceptive stimulus may be adequate, it is the balance and switch between exteroceptive and interoceptive information and the importance that they are assigned that could be different in ASD.

What would be the consequences of such dysfunctionning of the cingulo-opercular network salience network in ASD? On a speculative ground, it is possible that ASD subjects are more likely to focus on "internal sensations" than typical subjects. But paying too much attention to internal information can be a problem. Indeed, typical individuals can attend to higher level cognitive and social tasks by virtue of not needing to attend to the background delivery of interoceptive information. In ASD subjects, on the other hand, interoceptive information might become distracting, creating an imbalance between functional brain networks and sources of information. An interesting and challenging idea linked to the fact that the ASD physiopathology may be mainly expressed in changes in connectivity of core ROIs of the Cingulo-Opercular system such as AI and dACC is the recently proposed "Triple Network model" of psychopatology (Menon, 2011). This model proposes that most, if not all major psychopathologies are due to dysfunctions in large scale brain networks, principally involving these nodes of the CO system. The CO system dysfunction plays a major role: weak mapping from the AI - dACC areas gives rise to aberrant engagement of the FP system, compromising cognition and goal-relevant adaptive behaviour (Menon, 2011).

Finally, an open aspect of these results is whether the network changes between the two groups reveal a distinct pattern of thoughts, a different functional implementation of a comparable pattern of thoughts, or both. This note of caution is actually relevant for all studies comparing resting state activity between groups. The content of thoughts evoked during the resting state of a group of patients might be qualitatively different than those evoked by a typical group and this may account for the observed differences. By directing typicals and patients to different mental states and by observing opposed differences, the possibility that all the effects observed here reflect a different pattern of thought becomes unlikely but requires of course quantitative argumentation. Future work should elucidate whether indeed the observed imbalances in functional brain connectivity constitute a central aspect of ASD etiology, understand how they may relate to the organization of thought in the different states and validate their potential to become a clinically useful ASD biomarker.

**Acknowledgements**

This work is supported by the Human Frontiers Science Program. BW is supported by the CNRS. PB has a fellowship of the National Research Council of Argentina (CONICET) and was supported by a Human Frontiers Science Program research grant. The work is part of a collaborative ECOS project A08S03 between the French CNRS and the argentinian MinCyT.

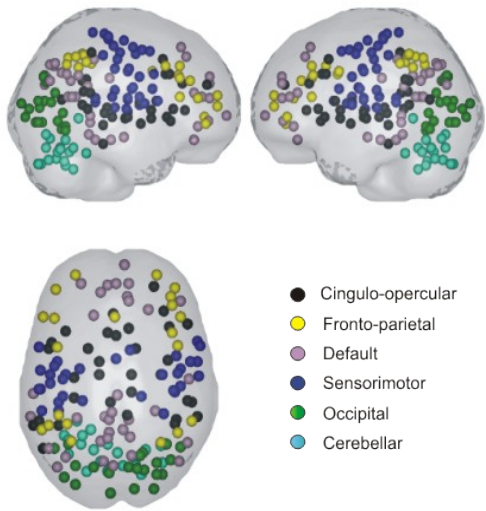

**Supplementary Figure 1:** 160 Regions of interest (ROIs) utilized in the analysis.

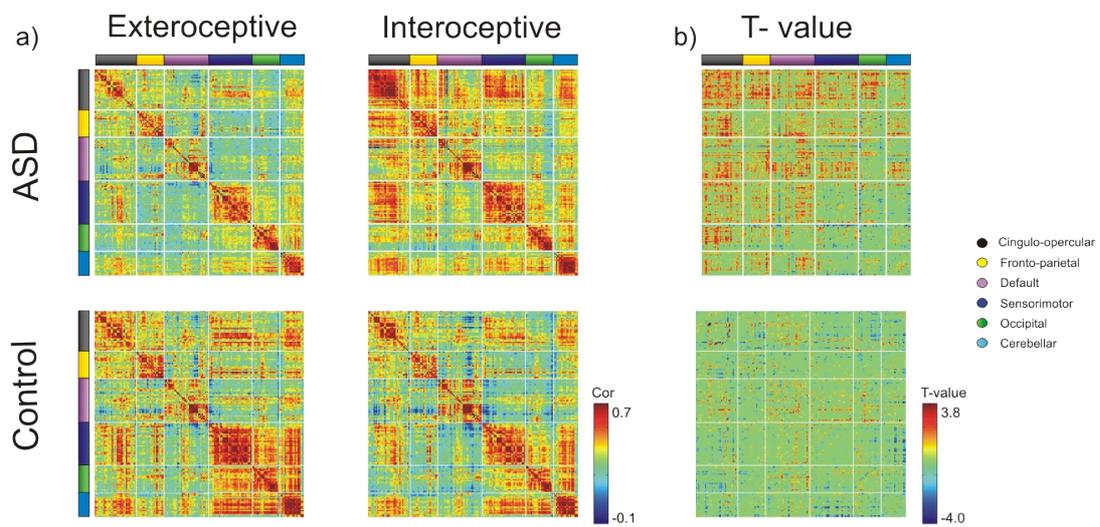

**Supplementary Figure 2:** Networks connectivity matrices. a) Averaged correlation matrices, for both groups in Exteroceptive and Interoceptive conditions. b) within group across conditions T-values distributions. Positive t-values (red) indicate higher connectivity in Interoceptive, negative t-values (blue) indicate higher connectivity in Exteroceptive.

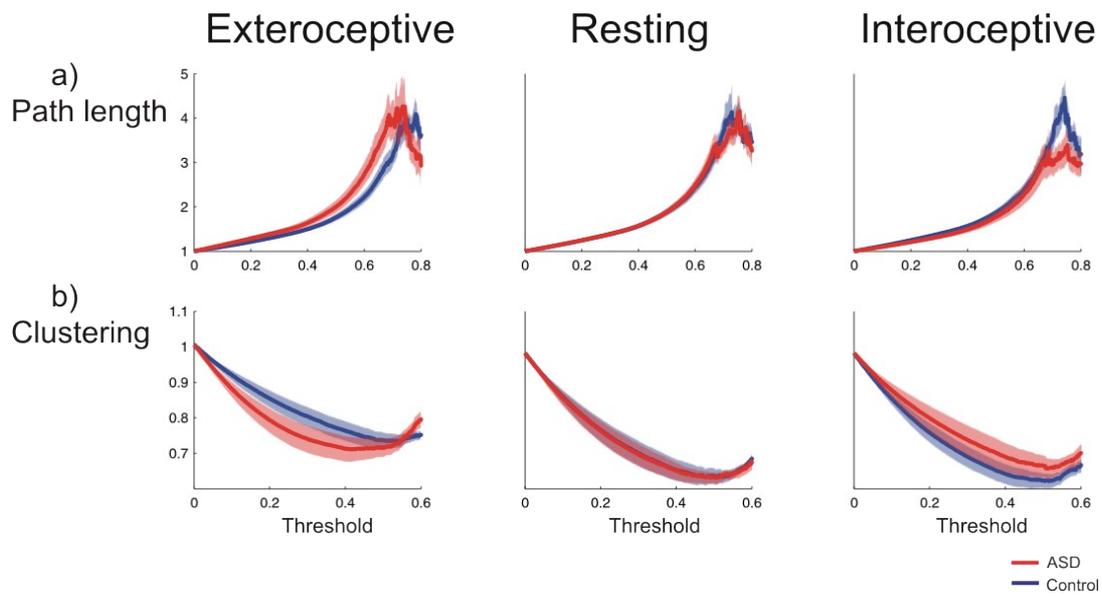

**Supplementary Figure 3:** Characteristic Path Length (a) and Clustering coefficient (b) for different thresholds (the cut-off to determine whether two nodes (ROIs) of the network are connected). Shadowed areas indicate mean standard error. Red (blue) curves show the dependence for the networks of the group of ASD (Typical) subjects.

| | Exteroceptive | |
|---|---|---|
| Source | F | p-value |
| Group effect | 15.55 | <0.001 |
| Type of Connection | 46.75 | <0.0001 |
| Interaction | 0.224 | 0.99 |
| | Resting | |
| Source | F | p-value |
| Group effect | 0.581 | 0.44 |
| Type of Connection | 42.89 | <0.0001 |
| Interaction | 0.34 | 0.99 |
| | Interoceptive | |
| Source | F | p-value |
| Group effect | 9.722 | <0.01 |
| Type of Connection | 37.73 | <0.0001 |
| Interaction | 0.48 | 0.96 |

**Supplementary Table 1.** Leave-one-out performance and p-value for all conditions and differences between conditions.

| Condition | Performance | P-value |
|---|---|---|
| Exteroceptive | 0,5 | >0,1 |
| Neutral | 0,625 | >0,1 |
| Interoceptive | 0,625 | >0,1 |
| Interoceptive – Exteroceptive | 0,916 | <0,0001 |
| Interoceptive - Neutral | 0,54 | >0,1 |
| Exteroceptive - Neutral | 0,45 | >0,1 |

**Supplementary Table 2.** a) Results for the ANOVA for connectivity values, for the three cognitive states.

| Mni coordinates | | | ROI label | Relative weighted degree difference | network |
|---|---|---|---|---|---|
| x | Y | z | | | |
| 38 | 21 | -1 | Anterior Insula | 1 | Cingulo-opercular |
| 9 | 20 | 34 | dACC | 0.94 | Cingulo-opercular |
| 27 | 49 | 26 | aPFC | 0.76 | Cingulo-opercular |
| -6 | 17 | 34 | Basal Ganglia | 0.74 | Cingulo-opercular |
| 32 | -59 | 41 | IPS | 0.70 | Fronto-parietal |
| 0 | 15 | 45 | mFC | 0.66 | Cingulo-opercular |
| 9 | -43 | 25 | Precuneus | 0.64 | Default |
| 51 | -30 | 5 | Temporal | 0.56 | Cingulo-opercular |
| -48 | 6 | 1 | vFC | 0.52 | Cingulo-opercular |
| -30 | -14 | 1 | Mid Insula | 0.50 | Cingulo-opercular |
| -2 | 30 | 27 | ACC | 0.47 | Cingulo-opercular |
| -61 | -41 | -2 | Inf Temporal | 0.47 | Default |
| 44 | 8 | 34 | dFC | 0.46 | Fronto-parietal |
| 42 | -46 | 21 | Sup temporal | 0.46 | Cingulo-opercular |
| 9 | 51 | 16 | vmPFC | 0.45 | Default |
| 40 | 17 | 40 | dFC | 0.40 | Fronto-parietal |
| 51 | -59 | 34 | Angular Gyrus | 0.39 | Default |
| -36 | 18 | 2 | Anterior Insula | 0.36 | Cingulo-opercular |
| 52 | -15 | -13 | Inf temporal | 0.36 | Default |
| 11 | -24 | 2 | Basal ganglia | 0.36 | Cingulo-opercular |
| 34 | 32 | 7 | vPFC | 0.35 | Cingulo-opercular |
| 9 | 39 | 20 | ACC | 0.27 | Default |
| 58 | -41 | 20 | Parietal | 0.27 | Cingulo-opercular |
| 44 | -52 | 47 | IPL | 0.27 | Fronto-parietal |
| -2 | -75 | 32 | Occipital | 0.26 | Default |
| 11 | -68 | 42 | Precuneus | 0.25 | Default |
| -30 | -28 | 9 | Post Insula | 0.25 | Cingulo-opercular |
| -59 | -47 | 11 | temporal | 0.25 | Cingulo-opercular |
| 8 | -40 | 50 | precuneus | 0.24 | Cingulo-opercular |
| 32 | -12 | 2 | Mid Insula | 0.24 | Cingulo-opercular |
| -5 | -52 | 17 | Post Cingulate | 0.23 | Default |
| 5 | -50 | 33 | Precuneus | 0.22 | Default |
| 10 | -55 | 17 | Post Cingulate | 0.21 | Default |
| -42 | 7 | 36 | dFC | 0.21 | Fronto-parietal |
| 46 | 28 | 31 | dlPFC | 0.21 | Fronto-parietal |
| 0 | 51 | 32 | mPFC | 0.21 | Default |
| -41 | -47 | 29 | Angular Gyrus | 0.20 | Cingulo-opercular |
| 28 | -37 | -15 | Fusiform | 0.20 | Default |
| 43 | -43 | 8 | Temporal | 0.20 | Cingulo-opercular |
| -46 | 10 | 14 | vFC | 0.19 | Cingulo-opercular |
| 29 | 57 | 18 | aPFC | 0.19 | Fronto-parietal |
| -52 | -63 | 15 | TJP | 0.18 | Cingulo-opercular |
| -12 | -12 | 6 | Thalamus | 0.17 | Cingulo-opercular |
| -9 | -72 | 41 | Occipital | 0.171 | Default |
| 40 | 36 | 29 | dlPFC | 0.16 | Fronto-parietal |
| -6 | -56 | 29 | Precuneus | 0.16 | Default |
| -59 | -25 | -15 | Inf Temporal | 0.15 | Default |
| -12 | -3 | 13 | Thalamus | 0.13 | Cingulo-opercular |
| -25 | 51 | 27 | aPFC | 0.13 | Default |
| -48 | -47 | 49 | IPL | 0.13 | Fronto-parietal |
| 1 | -26 | 31 | Post Cingulate | 0.12 | Default |
| -5 | -43 | 25 | Post Cingulate | 0.12 | Default |
| -29 | 57 | 10 | aPFC | 0.12 | Fronto-parietal |
| -3 | -38 | 45 | Precuneus | 0.12 | Default |
| -32 | -58 | 46 | IPS | 0.11 | Fronto-parietal |
| -11 | 45 | 17 | vmPFC | 0.10 | Default |
| -4 | -31 | -4 | Post Cingulate | 0.10 | Cingulo-opercular |
| 54 | -31 | -18 | Fusiform | 0.09 | Cingulo-opercular |
| -44 | 27 | 33 | dlPFC | 0.09 | Fronto-parietal |
| 46 | 39 | -15 | vlPFC | 0.09 | Default |
| -53 | -50 | 39 | IPL | 0.08 | Fronto-parietal |
| -41 | -40 | 42 | IPL | 0.08 | Fronto-parietal |
| 37 | -2 | -3 | Mid Insula | 0.07 | Cingulo-opercular |
| 39 | 42 | 16 | vlPFC | 0.07 | Fronto-parietal |
| 54 | -44 | 43 | IPL | 0.07 | Fronto-parietal |
| -52 | 28 | 17 | vPFC | 0.06 | Fronto-parietal |
| -8 | -41 | 3 | Post Cingulate | 0.06 | Default |
| 23 | 33 | 47 | Sup Frontal | 0.06 | Default |
| 6 | 64 | 3 | vmPFC | 0.05 | Default |
| -11 | -58 | 17 | Post Cingulate | 0.05 | Default |
| -1 | 28 | 40 | ACC | 0.05 | Fronto-parietal |
| -36 | -69 | 40 | IPS | 0.05 | Default |
| -48 | -63 | 35 | Angular Gyrus | 0.05 | Default |
| 11 | -12 | 6 | Thalamus | 0.04 | Cingulo-opercular |
| 8 | 42 | -5 | vmPFC | 0.04 | Default |
| -16 | 29 | 54 | Sup Frontal | 0.04 | Default |
| -35 | -46 | 48 | Post Parietal | 0.02 | Fronto-parietal |
| 45 | -72 | 29 | Occipital | 0.02 | Default |
| 14 | 6 | 7 | Basal Ganglia | 0.02 | Cingulo-opercular |
| -43 | 47 | 2 | vent aPFC | 0.02 | Fronto-parietal |
| -6 | 50 | -1 | vmPFC | 0.02 | Default |
| -42 | -76 | 26 | Occipital | 0.02 | Default |
| -20 | 6 | 7 | Basal Ganglia | 0.02 | Cingulo-opercular |
| 42 | 48 | -3 | vent aPFC | 0.01 | Fronto-parietal |
| -28 | -42 | -11 | Occipital | 0.00 | Default |
| 51 | 23 | 8 | vFP | 0.00 | Cingulo-opercular |
| -55 | -44 | 30 | Parietal | 0.00 | Cingulo-opercular |

**Supplementary Table 3.** Contribution of all ROIs to the prediction of ASD condition, relative to maximum.

**a) Cingulo-opercular network**

| Neutral condition | Number of voxels | Peak voxel z-score | Uncorrected p-value | P-value (FWE cluster corrected | Peak voxel MNI coordinates | Laterality | Label |
|---|---|---|---|---|---|---|---|
| | **Positive slopes** | | | | | | |
| | 242 | 4,43 | <0,001 | <0,001 | 28, -4, 56 | R | Middle frontal |
| | 211 | 4,08 | <0,001 | <0,001 | -36, -8, 48 | L | Precentral |
| | 155 | 3,80 | <0,001 | <0,001 | -14, -16, 50 | L | Cingulate Gyrus |
| | 90 | 3,56 | 0,001 | <0,001 | -10,-90,36 | L | Cuneus |
| | 117 | 3,47 | <0,001 | <0,001 | 6,-22,-36 | R | Pons |
| | 91 | 3.37 | 0,001 | <0,001 | 18,-84,0 | R | Calcarine |
| | **Negative slopes** | | | | | | |
| | 84 | 3,60 | 0,002 | <0,001 | -40,-80,-42 | L | Cerebellum |
| | 72 | 3,57 | 0,007 | <0,001 | 46,-68,46 | R | Angular |
| | 258 | 3,51 | <0,001 | <0,001 | -58,-60,-24 | L | Inferior Temporal |

| Interoceptive condition | Number of voxels | Peak voxel z-score | Uncorrected p-value | P-value (FWE cluster corrected | Peak voxel MNI coordinates | Laterality | Label |
|---|---|---|---|---|---|---|---|
| | **Positive slopes** | | | | | | |
| | 355 | 4,32 | <0,001 | <0,001 | -64, -26, 2 | L | Superior Temporal |
| | 248 | 4,06 | <0,001 | <0,001 | 60, -24, 34 | R | Supramarginal |
| | 262 | 3,86 | <0,001 | <0,001 | -6,2,56 | L | Cingulate Gyrus |
| | 151 | 4,01 | <0,001 | <0,001 | 26, 2, 56 | R | Middle frontal |
| | 127 | 3,39 | 0,002 | <0,001 | 60, 10,12 | R | Opercular |

**b) Fronto-parietal network**

| Neutral condition | Number of voxels | Peak voxel z-score | Uncorrected p-value | P-value (FWE cluster corrected | Peak voxel MNI coordinates | Laterality | Label |
|---|---|---|---|---|---|---|---|
| | colspan Negative slopes | | | | | | |
| | 999 | 4,38 | <0,001 | <0,001 | -20, -66, -16 | L | Occipital |
| | 386 | 4,36 | <0,001 | <0,001 | -34, -50, 66 | L | Parietal Superior |
| | 198 | 4,03 | <0,001 | <0,001 | 40, -74, -18 | R | Fusiform Gyrus |
| | 240 | 3,85 | <0,001 | <0,001 | 20,-68,-18 | R | Cerebellum |
| | 441 | 3,62 | <0,001 | <0,001 | 26,-96,10 | R | Occipital Superior |
| | 230 | 3.59 | <0,001 | <0,001 | 48,-56,-26 | R | Fusiform / Temporal Inferior |
| | 147 | 3.55 | 0,001 | <0,001 | 4,2,-14 | R | Anterior Cingulate / subcallosal gyrus |

| interoceptive condition | Number of voxels | Peak voxel z-score | Uncorrected p-value | P-value (FWE cluster corrected | Peak voxel MNI coordinates | Laterality | Label |
|---|---|---|---|---|---|---|---|
| | colspan Positive slopes | | | | | | |
| | 594 | 4,55 | <0,001 | <0,001 | 44, -54, 42 | R | Parietal Inferior |
| | 294 | 4,40 | <0,001 | <0,001 | 40, 14, 46 | R | Opercular cortex |
| | 353 | 4,23 | <0,001 | <0,001 | -48, -54, 52 | L | Parietal Inferior |
| | 454 | 4,00 | <0,001 | <0,001 | 12,-64,28 | R | Precuneus |
| | 164 | 3,92 | 0,001 | <0,001 | -6,-32,28 | L | Mid Cingulum |
| | 230 | 3.59 | <0,001 | <0,001 | 48,-56,-26 | R | Fusiform / Temporal Inferior |
| | 147 | 3.55 | 0,001 | <0,001 | 4,2,-14 | R | Anterior Cingulate / subcallosal gyrus |

c) **Default network**

| Neutral condition | Number of voxels | Peak voxel z-score | Uncorrected p-value | P-value (FWE cluster corrected | Peak voxel MNI coordinates | Laterality | Label |
|---|---|---|---|---|---|---|---|
| | colspan Negative slopes | | | | | | |
| | 518 | 3,78 | <0,001 | <0,001 | -22, -60, -16 | L | Cerebelum |
| | 124 | 4,09 | <0,001 | <0,001 | 24, -46, -48 | R | Cerebellum |
| | 176 | 3,69 | <0,001 | <0,001 | 12,-54,-14 | R | Cerebellum |
| | 230 | 3,65 | <0,001 | <0,001 | -30, -56, -48 | L | Cerebellum |
| | 311 | 3,59 | <0,001 | <0,001 | -22,-30,70 | L | Parietal Sup / Inf |
| | 92 | 3,53 | 0,002 | <0,001 | -10,4,16 | L | Caudate |
| | 84 | 3,38 | <0,001 | <0,001 | -34,-24,20 | L | Insula |
| | 210 | 3,37 | <0,001 | <0,001 | -44,-62,-6 | L | Temporal Inferior |
| | 116 | 3,37 | <0,001 | <0,001 | -32,-84,-30 | L | Cerebellum |
| | 78 | 3,28 | 0,008 | <0,001 | -34,-52-68 | L | Parietal Sup / Inf |
| | 84 | 3,25 | 0,005 | <0,001 | -12,50,-24 | L | Orbital cortex |
| | 93 | 3,22 | 0,002 | <0,001 | 34,-82,-14 | R | Occipital / Fusiform |
| | 82 | 3,19 | 0,006 | <0,001 | 30,-70,-36 | R | Cerebellum |
| | 145 | 3,17 | <0,001 | <0,001 | -22,-40,-28 | L | Cerebellum |
| | 93 | 3,13 | 0,002 | <0,001 | -8,28,-22 | L | Orbital cortex |

| Interoceptive condition | Number of voxels | Peak voxel z-score | Uncorrected p-value | P-value (FWE cluster corrected | Peak voxel MNI coordinates | Laterality | Label |
|---|---|---|---|---|---|---|---|
| | Positive slopes | | | | | | |
| | 216 | 3,93 | <0,001 | <0,001 | 2, -18, 2 | R | Thalamus |

**Supplementary Table 4.** Correlation between CO (a), FP (b), and DEF (c) and individual voxels, associated with ADOS score entered into multiple regression model (Whole brain corrected clusters (P < 0.001, extent threshold of 5).